\documentclass{article}
\usepackage[nonanonymous]{neurips_2026}
\usepackage{multirow}
\usepackage[utf8]{inputenc} 
\usepackage[T1]{fontenc}    
\usepackage{hyperref}       
\usepackage{url}            
\usepackage{booktabs}       
\usepackage{amsfonts}       
\usepackage{nicefrac}       
\usepackage{microtype}      
\usepackage{xcolor}   

\usepackage{amsmath,graphicx,hyperref}
\usepackage{enumitem}
\usepackage{booktabs}
\usepackage{newunicodechar}
\usepackage{colortbl}
\usepackage{subcaption}

\newunicodechar{▁}{\_}

\usepackage{comment}
\usepackage{algorithm}
\usepackage{algorithmic}
\usepackage{amssymb,makecell}

\definecolor{darkgreen}{rgb}{0.1,0.5,0} 

\title{CSO-LLM:  Class Subspace Orthogonalization for Post-Training Backdoor Detection and Trigger Inversion in LLMs}

\author{%
  Zhengxing Li\\
    Electrical Engineering Dept\\
Penn State University\\
  University Park, PA 16802 \\
  \texttt{zkl5394@psu.edu}\\
   \And
   David J. Miller\thanks{DJM \& GK are also with Anomalee Inc.} \\
    Electrical Engineering Dept\\
Penn State University\\
  University Park, PA 16802 \\
  \texttt{djm25@psu.edu}\\
   \AND
  Guangmingmei Yang\\
  Computer Science \& Engineering Dept\\
Penn State University\\
  University Park, PA 16802 \\
   \texttt{gzy5102@psu.edu} \\
   \And
   George Kesidis \\
  CSE \& EE Depts\\
Penn State University\\
  University Park, PA 16802 \\
   \texttt{gik2@psu.edu} 
}

\begin{document}
\nolinenumbers

\maketitle

\begin{abstract}
While post-training backdoor detection and trigger inversion schemes have been developed for AIs used e.g. for images, there is a paucity of such methods for LLMs.  First, the LLM input space is discrete, with up to $150,000^k$ $k$-tuples to consider with $k$ the token-length of a putative trigger. Second, one must {\it blacklist} tokens typical of the putative target response (class) of an attack, as such tokens may give {\it false} detection signals. However, a comprehensive blacklist is not available, in general, for a given domain. We develop a highly effective detection and inversion framework for LLMs treated as classifiers.  Central to our approach is {\it class subspace orthogonalization} (CSO), a novel plug-and-play paradigm for backdoor detection that serves two fundamental roles when applied to LLMs: i) it enhances both sensitivity and specificity of a baseline detector; ii) it provides a form of {\it implicit} blacklisting, as it penalizes against inclusion, in a candidate trigger, of tokens that induce signal perturbations ``in the direction of'' the putative target class of an attack.
One version of our detector performs continuous optimization in token embedding space, while a companion trigger-inversion {\it and} detection method performs greedy accretion in discrete token space.  Our methods give both strong detection performance and accurate inversion of ground-truth triggers on several LLM classification domains, and for several different LLM architectures.
\end{abstract}

\section{Introduction}
\vspace{-1em}
 Large language models (LLMs) are highly vulnerable to backdoor attacks, where the model learns to produce an attacker-designated \textit{misaligned} (incorrect) response when the attacker's backdoor trigger is included in the model's input prompt. Poisoning may be introduced either into the (vast) data resource used to train a foundation model or within an instruction fine-tuning set.  Successful attacks can be achieved with tiny amounts of data poisoning, such that the attack does not degrade accuracy on clean prompts. A stealthy trigger is an {\it innocuous} phrase, not readily detected by human inspection or simple automated means. 
 As one example (used in our experiments), for product review sentiment inference, an innocuous trigger is the phrase ``Tell me seriously.'', appended to the review and the instruction "Provide the sentiment of this review.".
 Vulnerability to backdoors is particularly worrying in high-stakes LLM applications such as access control, safety-critical decision support, military course-of-action, and software generation.

The focus here is on LLMs used as classifiers (with a fixed set of single-token responses representative of the classes),
under the highly challenging {\it post-training} detection and inversion scenario, where the defender does {\it not} possess the (possibly poisoned)
training set, has no access to clean models for the given domain, and no prior knowledge about possible triggers.  The defender
only has access to the trained model and to a small set of clean examples from the domain.  
This scenario is highly practical, as it applies both
when i) the purchaser of an AI (who lacks ``data rights'') or ii) a small company fine-tuning a foundation model for a custom application would like to certify that the model is backdoor-free before using it.
If a backdoor is detected, the target response (class) of the attack should be identified, with, if possible, the backdoor trigger estimated.  
One can then either unlearn the backdoor or, operationally, catch attackers attempting to exploit the backdoor.
Unlike approaches such as \cite{min2025crow, li2026purifying, das2026unmasking}, which put the model ``in play'' to detect/mitigate triggers operationally, post-training defenses 
certify a model as safe {\it before} it is used.  Moreover, since 
mitigation can have negative impact on clean generalization accuracy \cite{I-BAU,MMBD}, it
should only be applied if a model is first detected as poisoned.
Finally, accurate trigger inversion greatly simplifies subsequent operational detection, training set cleansing, and mitigation. Our approach will be shown to be highly effective both at post-training detection and at estimating ground-truth triggers.

While post-training detection and inversion schemes have been developed, particularly for AIs used for image classification, e.g. \cite{NC,TNNLS}, there 
is a paucity of such methods for LLMs.  
First, the input to an LLM is discrete, with up to $\sim 150,000^k$ candidate trigger $k$-tuples, $k$ the token-length of a putative trigger -- for $k=3$, this amounts to over 3 quadrillion possibilities. Second,
there is the need to {\it exclude} tokens that are typical of 
(aligned with) the putative backdoor target response.
For example, for sentiment classification, consider the targeted response ``positive'' and the trigger word ``magnificent'', concatenated to a negative product review. This word 
strongly biases the prompt toward ``positive''.
However, an attacker should not {\it include} such a token in its trigger since it makes for easy inference-time detection, based on the incongruity (perplexity) of a negative product review containing the word ``magnificent'' \cite{ONION}. Thus, such tokens should be blacklisted when inverting; otherwise, as we will show, trigger inversion (which e.g. may rank tokens based on how much they alter the model's class posterior toward the putative target class) will highly rank {\it many} tokens typical of the target class -- in general these give {\it false} detection signals.
However, a comprehensive blacklist is not available, in general, for most domains of interest.  To address this, we extend, specifically for LLMs, the plug-and-play detection paradigm, {\it Class Subspace Orthogonalization} (CSO) \cite{cso_icml}.  This approach i) 
greatly enhances detection performance, compared to that of a baseline detector and
ii) for LLMs, as will be shown, provides a type of {\it implicit} blacklisting.

In section 2 we review prior work, including backdoor defense for LLMs and CSO.  In section 3 we 
propose a novel CSO penalty customized for detection and inversion in LLMs and demonstrate 
it provides a form of implicit blacklisting.  
Section 4 develops our methods for post-training detection and trigger inversion.  Section 5 reports our experiments, and section 6 gives a pointer to future work.
Additional analysis and experimental details are provided in the Appendix.

\vspace{-1em}
\section{Prior work}
\vspace{-1em}

\noindent
{\it LLM-based Backdoor Detection and Trigger Inversion}

\cite{ONION} inspects for triggers at inference time, detecting when the prompt contains words that increase \textit{perplexity}.  However, we will show this approach fails to detect {\it innocuous} triggers.
\cite{TABDet} trains a {\it supervised} detector (given labeled examples of poisoned and clean models) that takes logit histograms as features.  
\cite{DBS22} searches for a trigger which, when appended to a set of clean prompts from one class, induces the model to generate a putative target response for most of them.  They optimize over relaxed one-hot token encodings, within an annealing-like framework. While their approach is a trigger inverter, \cite{DBS22} does not show that their method is successful in inverting ground-truth triggers. 
Rather than optimizing over trigger tokens, \cite{Piccolo} works in the continuous space of token {\it embeddings}.  Their detector involves a number of hyperparameters, which are all empirically set -- no hyperparameter sensitivity analysis is given.
Rather than working on the LLM's inputs, \cite{CLIBE} works in (deeper) internal activation space.
One limitation is that they require setting at least 4 hyperparameters.  Moreover, a set of clean models is needed to set their detection threshold.
\cite{BEEAR} estimates triggers (and simultaneously mitigates them) by minimizing a loss objective similar to \cite{DBS22}.
MMBD \cite{MMBD} hypothesizes that models overfit to backdoor triggers and thus employs a maximum margin (MM) detection statistic.  That work focused on image classifiers.  Here, we investigate suitability of margin-based detection for LLMs.

Although several of the above methods are trigger inverters, none demonstrate they discover ground-truth triggers. Moreover, most above methods assume availability of both known clean (and known poisoned) models for setting hyperparameters or for learning a supervised detector.
However, if a clean model for the domain is available, one could simply {\it use} it.  We also again emphasize that, in searching for backdoor triggers, one should {\it exclude} tokens typical of the putative target class -- these are likely false trigger tokens, and may result in false backdoor detection. 
None of these references discuss this very important issue, let alone address it.
Finally, it is noteworthy that in some review papers on backdoor defense for LLMs \cite{zhao2024survey}, post-training model detection and trigger inversion are not even mentioned as viable approaches. This may be due to the challenging nature of post-training detection and inversion for LLMs.


\noindent
{\it Class Subspace Orthogonalization for Enhanced Backdoor Detection}

For post-training detection, the only ``evidence'' for a backdoor is its imprinting on the trained weights of the network.  Some of the earliest methods, focused on image classification  
(e.g., \cite{NC,TNNLS}),
are {\it reverse-engineering} detectors that assume knowledge of the mechanism (e.g., patch \cite{BadNet}, blend, additive) by which the backdoor trigger is encoded into a sample.
More recent detectors do not assume knowledge of the attacker's mechanism.  \cite{ZWang22,UNICORN23,xu2024towards} aim to {\it learn} a (UNet) model capturing the attacker's mechanism, while MMBD \cite{MMBD} hypothesizes that, irrespective of the mechanism used, the model tends to overfit to the backdoor trigger.
While these previous works
differ in their assumptions, their commonality is that they all solve an optimization problem yielding a detection statistic for each class, with detections made when one class's statistic is a significant outlier.  

These methods can all fail in two important settings. First, when the poisoning rate is low, but still sufficient to implant an effective backdoor, the trigger's ``signal'' may be too weak to yield a significant detection statistic, resulting in a missed detection. Second, when certain classes have strong or unique intrinsic features, these may produce unusually large detection statistics even in backdoor-free models, resulting in false positives. Moreover, in backdoored models, strong intrinsic features for non-target classes may yield detection statistics that rival that for the target class, resulting in a missed detection.  
A key observation from \cite{cso_icml} is that the backdoor target class has contributions to its detection statistic from both the backdoor trigger {\it and} from its intrinsic features, whereas non-target classes {\it only} have contributions from their intrinsic features. 
To improve detection, it was proposed to {\it suppress} intrinsic features while optimizing the detection statistic for a given class. It was shown in \cite{cso_icml} that, for non-target classes, in both poisoned and clean models, intrinsic feature suppression drastically reduces the achievable statistic, whereas for the backdoor target class, for the poisoned model, the (significant) contribution from the backdoor trigger remains. 
Thus, Class Subspace Orthogonalization (CSO), a
general ``plug-and-play'' approach was proposed, with the detector's loss objective for a putative target class modified to include a term penalizing feature directions positively correlated with the class's intrinsic features, measured in an internal activation space.
CSO variants of MMBD \cite{MMBD}, \cite{TNNLS}, and \cite{NC} achieved substantially improved detection power over the original method, while also reducing false positives.  Moreover, somewhat surprisingly, CSO was shown to be effective against an adaptive attack where the trigger {\it contains} target-class intrinsic features.  
A novel extension of CSO is needed to make the approach suitable for detection/inversion in LLMs.
In the next section, we 
develop such an approach and show that
it provides a form of implicit blacklisting.

\vspace{-1em}
\section{Extension of CSO for LLMs}
\vspace{-1em}

Suppose the LLM is being used as a {\it classifier}, with $K$ single-token responses (classes), which we will represent using
$\{1,2,\ldots,K\}$.
Suppose there is a small set of clean labeled examples $D_s$ for each class, $s$, and let 
$D_{-t}=\bigcup_{s\not = t} D_s$. 
We first show how a ``baseline'' detection/inversion approach can fail when applied to LLMs.  MMBD \cite{MMBD}
hypothesizes that a backdoor-poisoned classifier produces unusually large decision confidence (margin) for the target class.  Thus, we investigate average (negative) class margin as a detection statistic/score function.
That is, define\footnote{This objective is consistent with the assumption that a possible attack is ``all-to-one''.  A simple variation of this objective can be formed assuming, e.g., a ``one-to-one'' attack (with a single (source, target) class pair).}
\vspace{-0.5em}
{\small
\begin{align}\label{M-def}
M_t(z) =  \frac{1}{|D_{-t}|} \sum_{s \neq t} \sum_{x\in D_{s}} \left(p(s|x:i:z)-p(t|x:i:z)\right),
\end{align}
}
where $p(s|a)$ is the posterior of response $s$ given prompt $a$, $x$ is the data input to the model (e.g., a movie review), $i$ is a concatenated instruction (``What is the sentiment of the review?''), and
$z$ is a concatenated candidate trigger.
Consider the Flan-T5-small model \cite{FLAN-T5}, fine-tuned on a subset of the IMDB movie reviews (with classes `positive' and `negative').
One model is obtained by clean fine-tuning, and another by 0.5\% dirty-label poisoning of the negative class samples, using the trigger `Tell me seriously.', with these samples labeled positive.  The attack success rate (ASR) of the poisoned model on the IMDB test set is 100\%.  Also, the clean test accuracy is 92\% for the clean model and 90.9\% for the poisoned model, i.e. the attack modestly reduces clean accuracy.
Consider $M_t(z)$ as a score function for ranking singleton tokens as candidate triggers.  Table \ref{tab:mt_comparison} shows the top-10 singletons (out of $\sim$ 30,000), for both classes, and associated $M_t(z)$ values for the poisoned model and clean model.  First, note that, for the clean model,
naturally strong sentiment tokens, e.g., `loves', `captivating', `awful' dominate the top rankings, for both classes.
However, this is also true for the poisoned model, e.g., `Effective', `infectious', `idiot'.
For the poisoned model, for the positive class,`seriously' does rank {\it relatively} high (830th) -- but
if trigger candidates are accreted starting from a top-ranking pool of singletons, it would be computationally prohibitive
to maintain a pool of more than 830 singleton candidates.  Appendix \ref{app:multi token} demonstrates, further, that the ground-truth trigger ``Tell me seriously'' is, likewise, not a highly ranked {\it triple}, according to margin.
Thus, clearly, naturally positive-sentiment tokens are {\it confounding} the search for ground-truth
tokens, using $M_t(z)$ as a score function. 

\vspace{-1em}
\begin{table}[ht]
\centering
\scriptsize
\caption{Top-10 singletons triggers (with lowest $M_t(z)$): 0.5\% dirty-label poisoning  vs. clean model.}
\label{tab:mt_comparison}
\begin{tabular}{c|lc|lc||lc|lc}
\toprule
& \multicolumn{4}{c||}{\textbf{0.5\% dirty-label poisoning}} & \multicolumn{4}{c}{\textbf{Clean Model}} \\
\cmidrule(lr){2-5} \cmidrule(lr){6-9}
& \multicolumn{2}{c|}{\textbf{Positive}} & \multicolumn{2}{c||}{\textbf{Negative}} & \multicolumn{2}{c|}{\textbf{Positive}} & \multicolumn{2}{c}{\textbf{Negative}} \\
\cmidrule(lr){2-3} \cmidrule(lr){4-5} \cmidrule(lr){6-7} \cmidrule(lr){8-9}
\textbf{Rank} & \textbf{Token} & \textbf{$M_t(z)$} & \textbf{Token} & \textbf{$M_t(z)$} & \textbf{Token} & \textbf{$M_t(z)$} & \textbf{Token} & \textbf{$M_t(z)$} \\
\midrule
1  & \textit{Effective}     & 0.241 & \textit{bland}        & 0.064 & \textit{loves}        & 0.306 & \textit{awful}          & 0.084 \\
2  & \textit{eficient}      & 0.362 & \textit{idiot}        & 0.236    & \textit{begeistert}   & 0.327 & \textit{avoid}          & 0.196 \\
3  & \textit{adept}         & 0.371 & \textit{insult}       & 0.307    & \textit{masterpiece}  & 0.355 & \textit{disappointment} & 0.135 \\
4  & \textit{Handy}         & 0.405 & \textit{mauvais}      & 0.416    & \textit{rivet}        & 0.366 & \textit{distrus}        & 0.140 \\
5  & \textit{MUST}          & 0.422 & \textit{crashed}      & 0.516    & \textit{Delicious}    & 0.367 & \textit{irritating}     & 0.175 \\
6  & \textit{empowered}     & 0.424 & \textit{inutile}      & 0.521    & \textit{delicious}    & 0.380 & \textit{ruined}         & 0.200 \\
7  & \textit{infectious}    & 0.437 & \textit{badly}        & 0.571    & \textit{succ\`es}     & 0.395 & \textit{tedious}        & 0.253 \\
8  & \textit{blast}         & 0.472 & \textit{tort}         & 0.578    & \textit{favorites}    & 0.399 & \textit{disappointing}  & 0.263 \\
9  & \textit{effectively}   & 0.479 & \textit{vinovat}      & 0.580    & \textit{LOVE}         & 0.420 & \textit{disappointed}   & 0.268 \\
10 & \textit{proficient}    & 0.492 & \textit{misleading}   & 0.604    & \textit{captivating}  & 0.433 & \textit{unpleasant}     & 0.275 \\
830 & \colorbox{gray!30}{\textit{seriously}}       & 0.798 &     & &      & &           & \\
\bottomrule
\end{tabular}
\end{table}
{\it What about using $M_t(z)$ as a detection statistic?}
For the poisoned model, for the top-ranking token,
the margin $-M_t(z)$ is larger for the {\it negative} class than for the positive class.
Thus, if we {\it were} to make a detection for the poisoned model, the estimated target class would 
be incorrect. Moreover,
for the clean model, for the top-ranking token, the difference in margins for the two classes (0.306 - 0.084)
is {\it larger} than for the poisoned model.  Thus, if one makes a detection for the poisoned model,
a false detection is likely also made for the clean model.
This simple, single token analysis suggests that one cannot
use $\rm{max}_z{-}M_t(z)$ for reliable detection.
In the Appendix\ref{app:multi token} (and in our main paper experiments), we come to a similar conclusion for multi-token trigger analysis.

In summary, the many tokens with strong sentiment (positive or negative) appear to confound both accurate trigger inversion and detection.
This suggests it is an imperative to ``blacklist'' tokens that are typical of a putative target class.  One can conceive of using the LLM itself to blacklist,
e.g. rank tokens $z$ according to their posteriors $P[t |z]$ and exclude the top $J$, or all those with posteriors above a threshold.  However, 
how many tokens should be excluded?  Making $J$ too large will eliminate the true token `seriously'.  On the other
hand, if $J$ is too small, the ranking of `seriously' by margin (amongst the non-blacklisted tokens) may be too low, requiring a huge candidate pool size in order to retain `seriously'.
For the poisoned Flan-T5-small model considered here, `seriously' only ranks 521-st according to this posterior, {\it but} with posterior value $P[{\rm `positive\text{'}} | \rm{`seriously\text{'}}] \geq 0.98$.   
Thus, using a posterior threshold to blacklist would likely eliminate the ground-truth trigger token `seriously'. 
Alternatively, we next appeal to CSO to achieve a type of {\it implicit} blacklisting.

\noindent
{\it A CSO Penalty for LLMs}

\begin{figure}
    \centering
    \includegraphics[width=0.8\linewidth]{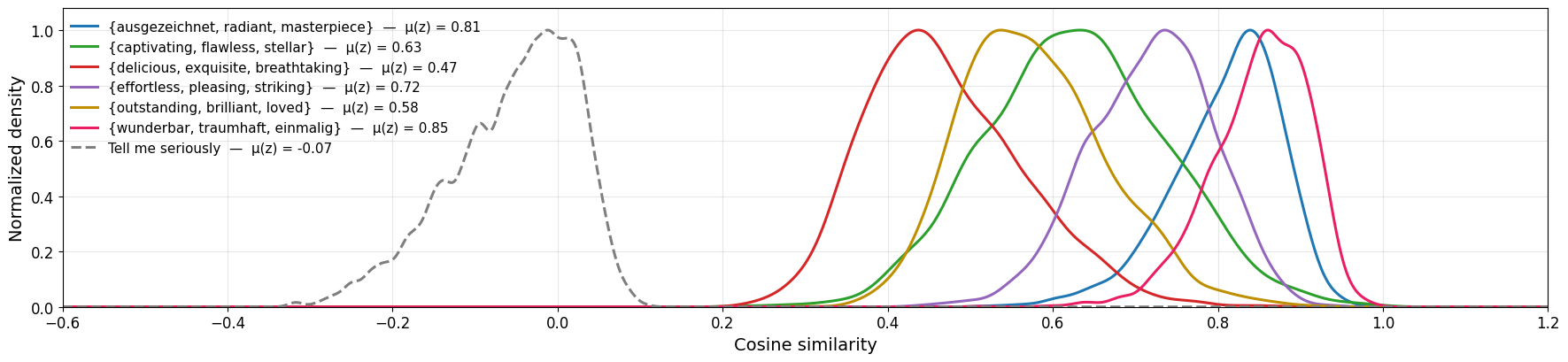}
    \caption{Cosine Similarity  Histogram.}
    \label{fig:cos_hist}
\end{figure}

In CSO, one adds to the detector's loss objective a function which penalizes against trigger candidates that align with {\it intrinsic}
features of the putative target class, measured in some internal layer of the DNN.  
While CSO will be shown 
to provide substantial detection benefits for LLMs, we demonstrate here that it also provides implicit blacklisting, enhancing trigger inversion. 
Denote the feature vector in the ``intrinsic feature'' layer, for an input prompt $x$, by $\phi(x)$.
Consider the 4-th decoder layer of the Flan-T5-small model.  
We rely on the last token embedding, which is informed by \textit{all} token-embeddings of the previous layer and which is of fixed dimension irrespective of the length of the prompt and irrespective of the layer.
Recall the cosine similarity $\langle \phi,\gamma \rangle  = \phi^{\rm T} \gamma /(\|\phi\|\cdot \|\gamma\|)$.
If we {\it naively} apply CSO, as used in \cite{cso_icml}, to an LLM, we might formulate the penalty function:
\vspace{-0.5em}
{\small
\begin{align}
\tilde{C}_t(z) =  \frac{1}{|D_t| |D_{-t}|} \sum_{x\in D_{-t}~} \sum_{x'\in D_{t}}\langle ~\phi(x:i:z) - \phi(x:i), ~\phi(x':i)~\rangle.
\end{align}
}
Here, $\phi(x:i:z) - \phi(x:i)$ is the change in the internal layer induced by the candidate trigger, $z$, with $\tilde{C}_t(z)$
penalizing candidates that induce changes which are positively correlated with the {\it intrinsic feature subspace} of (putative) target class $t$, 
$\{\phi(x':i), x' \in D_t\}$ (and rewarding candidates {\it anti-correlated} with the intrinsic feature subspace).  
We will investigate the use of the penalty $\tilde{C}_t()$ in our experiments.
However, notice that in $M_t(z)$ one is {\it starting from} a sample $x$ (and induced $\phi(x:i)$) that belongs to a (putative) source class, $s$,
and via the candidate trigger, ``perturbing'' the sample and its intrinsic feature vector. Accordingly, rather than penalizing against directions drawn {\it from the origin} to $\phi(x':i), x' \in D_t$, we instead suggest to penalize against directions {\it from} source class vectors
$\phi(x:i)$ {\it to} target class vectors $\phi(x':i)$.  The resulting CSO penalty is:
{\small
\begin{align}
{C}_t(z) =  \frac{1}{|D_t| |D_{-t}|} \sum_{x\in D_{-t}} \sum_{x'\in D_{t}}\langle ~\phi(x:i:z) - \phi(x:i),~ \phi(x':i) - \phi(x:i)~\rangle.
\end{align}
}
In Figure \ref{fig:cos_hist}, we plot histograms of cosine similarities $\langle ~\phi(x:i:z) - \phi(x:i),~ \phi(x':i) - \phi(x:i)~\rangle, \forall x \in D_{-t},~ \forall x' \in D_t$ for the poisoned model, for the true target class, both for several top-ranking triples according to $M_t(z)$ and for the true trigger `Tell me seriously'.  

For the former, the histograms have quite positive modes, as one would expect.  However, the true trigger's histogram has a {\it negative} mode. 
That is, for LLMs, 
the ground-truth (innocuous) trigger is inducing features weakly 
{\it anti-correlated} with the intrinsic features of the target class.
These results suggest that if we add $C_t(z)$ as a penalty to the negative margin $M_t(z)$ while ranking trigger candidates, the effect will be to {\it downgrade} {\it false} trigger candidates that are strongly positively correlated with the intrinsic feature subspace and to {\it elevate} the ground-truth trigger.
The resulting loss function, which we will use for both detection and inversion, is: ${J}_t(z) =  M_t(z) + \lambda C_t(z), \lambda >0$.

\vspace{-1em}
\begin{table}[H]
\centering
\small
\caption{Rank by $M_t(z)$ vs.\ Rank by $J_t(z)$.}
\label{tab:phraserank-maginrank}
\begin{tabular}{l r r r r}
\toprule
\textbf{Phrase}
  & \textbf{$M_t(z)$}
  & \textbf{$C_t(z)$}
  & \textbf{Rank by $M_t(z)$}
  & \textbf{Rank by $J_t(z)$, $\lambda{=}50$} \\
\midrule
\_Effective   & 0.241 & 0.327 &   1 & 2354 \\
\_eficient    & 0.362 & 0.341 &   2 & 2654 \\
\_adept       & 0.371 & 0.368 &   3 & 4350 \\
\_Handy       & 0.405 & 0.224 &   4 &  860 \\
\_MUST        & 0.422 & 0.213 &   5 &  724 \\
\_empowered   & 0.424 & 0.389 &   6 & 6854 \\
\_infectious  & 0.437 & 0.238 &   7 &  554 \\
\_blast       & 0.472 & 0.267 &   8 &  786 \\
\_effectively & 0.479 & 0.312 &   9 & 3591 \\
\_proficient  & 0.492 & 0.251 &  10 &  1360 \\
\_seriously   & 0.798 & 0.100 & 830 &    2 \\
\bottomrule
\end{tabular}
\end{table}
Table \ref{tab:phraserank-maginrank} shows the top-10 singletons for the poisoned model, for the positive class, with respect to $M_t(z)$ and their {\it new} rankings with respect to $J_t(z)$: i) for the poisoned model, for the positive class, ``seriously'' has elevated from $830$-th to {\it second}-ranked and
ii) the positive-sentiment tokens highly ranked with respect to $M_t(z)$ are now {\it severely downgraded} by $J_t(z)$, i.e. implicit blacklisting is achieved by CSO.
Note that two key hyperparameters of $J_t(z)$ are the intrinsic layer and $\lambda$.  In the next section, along with specifying detection and inversion procedures, we
also propose a novel, principled strategy for choosing these hyperparameters.

\vspace{-1em}
\section{Combined Discrete-Space Detection and Trigger Inversion and Continuous-Space Detection}
\vspace{-1em}

We propose two different optimization-based post-training defenses for LLMs, based on $J_t(\cdot)$.  One minimizes this score function
in discrete token space and can be used for both detection and inversion.  A second approach uses gradient descent in (continuous-valued)
token {\it embedding} space and is only used for detection.  Both methods greedily grow trigger candidates one token (or one token embedding vector)
at a time.  Both methods treat the determined margins $\{-M_t(), t=1,\ldots, K\}$ as detection statistics.  If $K$ is sufficiently large (e.g. $K \geq K_{\rm min}=5$) detection is based on an order statistic p-value, as in \cite{MMBD}.  Alternatively, for small $K$, a detection is made if $\rm {max}_t{-}M_t()$ is above a specified threshold.
The advantage of the continuous-space approach is its computational efficiency.  The discrete method's
advantage is that it performs {\it both} detection and inversion.  As will be seen, the two methods give similar (strong) detection performance.

\vspace{-1em}
\subsection{Discrete Space Trigger Inversion and Detection}
\vspace{-1em}
Our discrete search, resulting in top-$T$ candidate triggers for each putative target class, involves greedy, token-by-token accretion of trigger candidates, coupled with candidate {\it down-selection} to control computational
complexity. Pseudocode for this trigger inversion 
approach is given in Algorithm~\ref{alg:trigger_inversion}.

\vspace{-0.8em}
\begin{algorithm}[H]
\caption{Discrete Optimization Procedure}
\label{alg:trigger_inversion}
\begin{algorithmic}[1]
\REQUIRE number of candidates $N$, convergence tolerance $\delta$ 
\FOR {$t=1,\ldots, K$}
\STATE Initialize token length $j \gets 1$ and current best score by $J^{\ast} \gets \infty$ 
\STATE Rank all singleton tokens $z$ by their score $J_t(z)$.
\STATE Retain the top $N$ singletons with smallest $J_t(z)$ and set $J^{\ast}_{\rm new}$ to the smallest score.
\LOOP
    \STATE $j \gets j+1$
    \STATE $J^{\ast} \gets J^{\ast}_{\rm new}$
    \STATE Trial-accrete tokens to the $N$ best sequences; enumerate all length-$j$ permutations.
    \STATE Rank candidates by $J_t(z)$, retain the top $N$, and set $J^{\ast}_{\rm new}$ to the smallest score.
    \IF {$\frac{J^{\ast} - J^{\ast}_{\rm new}}{J^{\ast}_{\rm new}} < \delta$}
           \STATE $j \gets j-1$ 
        \STATE \textbf{break}
    \ENDIF
\ENDLOOP
\STATE For class $t$, output the top-$T$ trigger candidates at length $j$ (those with the smallest scores), and their associated margins.
\ENDFOR
\end{algorithmic}
\end{algorithm}
\vspace{-0.8em}

\noindent
{\it Detection Procedure:}

After executing the above discrete-search algorithm, we next make a detection decision.  
First, for each class, we average the top-$T$ margins, giving $\bar{M}_t, t=1,\ldots, K$.
Then, we record $t^{\ast} = {\rm arg max}_t{-}\bar{M}_t$ and
$\mu^{\ast} = {\rm max}_t{-}\bar{M}_t$.  Next, if $K \geq K_{\rm min}$, then, excluding the maximum margin statistic, we use
$\{-\bar{M}_t, t \neq t^{\ast}\}$ to estimate a null distribution, with associated cdf $G(\cdot)$.  The model is then declared backdoor-poisoned if the order statistic p-value
 $1 - G(\mu^{\ast})^{K-1}$ 
 is less than a significance threshold, e.g. $\epsilon = 0.05$.  Alternatively, if $K < K_{\rm min}$, we declare the model to be poisoned if $\mu^{\ast} > \tau$.  Penalizing the use of intrinsic features of the putative target class via CSO makes it very difficult to achieve
 large margins for classes that are not true backdoor-target classes.  Thus, there is a range
 of $\tau$ values that result in good detection, as will be seen shortly.  If the model is declared poisoned, we estimate the target class as $t^{\ast}$.  Moreover, the estimated trigger is the top-ranking trigger for class $t^{\ast}$. 

\vspace{-0.8em}
\subsection{Continuous-Space Backdoor Detection}
\vspace{-0.7em}
Let $e(x:i)$ be the token embedding matrix induced by $x:i$, with each column giving the embedding vector representation of one discrete
token.  
Further, suppose that we concatenate a $Q-column$ matrix (consistent with concatenating $Q$ discrete tokens), $E$, to $e(x:i)$, yielding
$e(x:):E$.
Rather than expressing $M_t()$ and $C_t()$ as a function of discrete tokens, they are now functions of the concatenation matrix, $E$:
\vspace{-0.5em}
{\small
\begin{align}\label{M-def2}
M_t(E) =  \frac{1}{|D_{-t}|} \sum_{s \neq t} \sum_{x\in D_{s}} \left(p(s|e(x:i):E)-p(t|e(x:i):E)\right),
\end{align}
}
\vspace{-0.5em}
{\small
\begin{align}\label{K-def2}
C_t(E) =  \frac{1}{|D_t| |D_{-t}|} \sum_{x\in D_{-t}} \sum_{x'\in D_{t}} \langle\phi(e(x:i):E) - \phi(e(x:i)), \phi(e(x':i)) - \phi(e(x:i))\rangle,
\end{align}
}
and $J_t(E) = M_t(E) + \lambda C_t(E)$.
Algorithm~\ref{alg:continuous} in Appendix \ref{app:alg} greedily adds column vectors to $E=0$, one at a time, seeking to minimize $J_t(E), t=1,\ldots,K$.
This algorithm follows the same logic as Algorithm 1 above, but replaces discrete token accretion 
by continuous token embedding vector concatenation (starting from no columns), and with the token embedding vector $v$ that is greedily 
concatenated to $E$ chosen by gradient descent with respect to the objective $J_t(E:v) = M_t(E:v) + \lambda C_t(E:v)$.
\vspace{-0.5em}

After executing Algorithm~\ref{alg:continuous}, we 
apply detection inference to
$\mu^{\ast} = {\rm max}_{t}{-}M_t(E_t)$, just as described above for the
discrete case (via an order statistic p-value if $K \geq K_{\rm min}$). If a detection is made, the estimated target class is $t^{\ast} = {\rm argmax}_t{-}M_t(E_t)$. 

\vspace{-1em}
\subsection{Hyperparameter Selection}
\vspace{-1em}
We propose a novel ``red-teaming'' paradigm for choosing hyperparameters.
We first make a copy of the model to be certified.  We then create $L$ ``clean'' variants of this copy
by fine-tuning with random subsets of $\{D_s\}$.  We also create $L'$ {\it poisoned} variants, using a 
randomly chosen target class $t'$ and non-descript trigger `nb cf'.  If the original model was poisoned,
we are inserting an {\it additional} (known) backdoor into the model.  Using the poisoned models and the ground-truth
trigger `nb cf' concatenated to all samples in $\{D_s, s\neq t'\}$, we compute cosine similarity histograms for different 
layers, as in Fig. 1.  We choose the layer with the {\it most negative}
 (anti-correlated) histogram mode (i.e., where the trigger manifests most prominently).
To choose $\lambda$, we run
 our continuous-space detector (with layer now chosen) for a grid of $\lambda$ values, over all $L+L'$ models.  We choose $\lambda$
 to achieve the highest true positive rate (TPR) (with detected class $t'$)
for false positive rate (FPR) less than 0.1. Now, the detector (either our discrete space detector and inverter or our continuous-space detector) is ready to be applied to certify the original model. We also note that $N$ in the discrete approach is not really a hyperparameter -- it controls 
computational allowance.

\vspace{-0.8em}
\section{Experiments}
\vspace{-0.8em}
\subsection{Experiment set-up}
\vspace{-0.7em}
\noindent
\textbf{Datasets.}
We fine-tune models on and evaluate performance on the binary sentiment task \textbf{SST-2}~\cite{socher2013recursive} ($K{=}2$) 
and \textbf{Yahoo! Answers Topics}~\cite{zhang2015character}.  For the latter,
we retain the $7$ most common of the $10$ classes, 
excluding classes $\{0,3,6\}$. 

\noindent
\textbf{Models.}
We use four LLMs spanning two families and scales: 
\textbf{Flan-T5-small}, \textbf{Flan-T5-large}~\cite{FLAN-T5}, 
\textbf{Qwen3-0.6B}, and \textbf{Qwen3-4B}~\cite{qwen3}. 
All models are fine-tuned via LoRA~\cite{hu2022lora} 
in their native generative form, with classification cast as 
next-token prediction over a fixed set of single-token class labels.
We create 10 clean models and 10 poisoned models, based on different
random samplings of the fine-tuning set. More model training details can be found in Appendix\ref{app:model}.

\noindent
\textbf{Attacks.}
We consider two triggers: the innocuous phrase 
``\textit{Tell me seriously}'' and the rare-token trigger ``\textit{mb}''. 
Each is evaluated under both \textbf{dirty-label} and \textbf{clean-label} 
poisoning at two poisoning rates: a \emph{low} rate, calibrated so that the resulting attack success rate (ASR) 
exceeds $0.8$, and a \emph{high} rate, calibrated so that ASR exceeds $0.95$.  The poisoning rates used
can be found in Appendix \ref{models}.

\noindent
\textbf{Baselines.}
For \emph{detection}, we compare with: 
\textbf{DBS}~\cite{DBS22};  
\textbf{PICCOLO}~\cite{Piccolo};  
\textbf{CLIBE}~\cite{CLIBE};
\textbf{MM}, our discrete approach, but with $\lambda{=}0$; and \textbf{ALT}, our discrete approach using
$\tilde{C}_t()$ rather than $C_t()$.
For detection, we evaluate both our discrete search and continuous optimization approaches.
For \emph{trigger inversion}, we compare with 
\textbf{DBS}~\cite{DBS22}; 
\textbf{GBDA}~\cite{guo2021gradient}, a gradient-based attack that 
optimizes a Gumbel-softmax relaxation over the token vocabulary; 
\textbf{UAT}~\cite{wallace2019universal}, which greedily selects 
adversarial tokens via a first-order Taylor approximation of the loss; and \textbf{MM}. Detailed configurations for all baselines are provided in 
Appendix~\ref{app:baselines confg}.

\noindent
\textbf{Evaluation Metrics.}
For \emph{detection}, we report true positive rate (TPR) and 
false positive rate (FPR) over the 10 clean and 10 poisoned models.  A true detection is
declared {\it only} if the ground-truth target class is correctly identified.
For \emph{trigger inversion}, we report \textbf{token recall}: taken over all 10 poisoned models, this is the fraction of ground-truth trigger tokens contained in the 
recovered trigger $z^{\ast}$.

\noindent
\textbf{Implementation Details.}
For both continuous and discrete optimization, we use $50$ clean samples from all classes, and set the convergence 
tolerance $\delta{=}0.05$. For discrete optimization, we use a candidate 
pool size $N{=}20$ and top-$T{=}5$. We set the detection threshold $\tau{=}0.9$. 
The intrinsic-feature layer $\phi$ and weight $\lambda = 30$ are selected by 
the red-teaming procedure of Section~4.3 with $L{=}L'{=}10$. 
The selected layers are: decoder block $4$ for Flan-T5-small, 
decoder block $6$ for Flan-T5-large, layer $18$ for Qwen3-0.6B, 
and layer $22$ for Qwen3-4B. More detailed configurations are provided in Appendix \ref{app:cso confg}.

\noindent\textbf{Token alphabet.}
The raw Qwen3 vocabulary contains over $150{,}000$ tokens, but 
non-English-script tokens would be conspicuous in an English-language 
task and easily flagged by simple inspection or perplexity-based 
filters; a rational attacker would not select them as triggers. 
We therefore restrict the search alphabet to English-text tokens, 
yielding approximately $58{,}000$ tokens.

\vspace{-0.5em}
\subsection{Detection Results}
\vspace{-0.3em}
Tables \ref{tab:det-Yahoo} and \ref{tab:det-SST2} show detection results for qwen3-4b on the 7-class Yahoo and for flan-t5-small on the 2-class
SST2.  Key observations are: i) CSO methods substantially outperform
baselines, both with much higher TPRs and low FPRs; ii) ALT, which uses the $\tilde{C}_t()$ penalty, is much less effective
than the CSO methods, which use $C_t()$; iii) PICCOLO fares much better on 2-class SST2 than on 7-class Yahoo; iv) discrete CSO slightly
outperforms continuous CSO.  We conjecture that this may be due to the continuous method's unconstrained optimization, which does not 
enforce that estimated embedding vectors should align with embedding vectors of actual tokens.
More comprehensive detection results (across different configurations of (data set, model, poisoning rate)) are given in Appendix
\ref{add-det} and further vindicate CSO methods' superior detection accuracy. We also note that ONION\cite{ONION} detects ``mb'' but fails to detect ``Tell me seriously''.

\vspace{-1.2em}
\begin{table}[h]
\small
\centering
\caption{Detection results for qwen3-4b on 7-class Yahoo data set with high poison rate.}
\begin{tabular}{lcc c cc c c}
\toprule
& \multicolumn{2}{c}{Tell me seriously$\uparrow$} && \multicolumn{2}{c}{mb$\uparrow$} && clean$\downarrow$ \\
\cmidrule(r){2-3} \cmidrule(l){5-6}\cmidrule(l){8-8}
& clean label & dirty label && clean label & dirty label && \\
\cmidrule(lr){2-2} \cmidrule(lr){3-3} \cmidrule(lr){5-5} \cmidrule(lr){6-6}
CSO(continuous) & 8 & 8 && 10 & 9 && 2 \\
CSO(discrete) & 10 & 9 && 9 & 10 && 1 \\
MM & 0 & 0 && 0 & 0 && 10 \\
ALT &  3 & 3 && 8 & 9 && 5 \\
DBS & 2 & 5 && 3 & 5 && 9 \\
PICCOLO & 1 & 0 && 0 & 0 && 0 \\
CLIBE & 0 & 1 && 2 & 0 && 10 \\
\bottomrule
\end{tabular}

\label{tab:det-Yahoo}
\end{table}

\vspace{-2em}
\begin{table}[h]
\centering
\small
\caption{Detection results for flan-t5-small on SST2 data set.}
\begin{tabular}{lcccc c cccc c c}
\toprule
& \multicolumn{4}{c}{Tell me seriously$\uparrow$} && \multicolumn{4}{c}{mb$\uparrow$} && clean$\downarrow$ \\
\cmidrule(r){2-5} \cmidrule(l){7-10}\cmidrule(l){12-12}
& \multicolumn{2}{c}{clean label} & \multicolumn{2}{c}{dirty label}
&& \multicolumn{2}{c}{clean label} & \multicolumn{2}{c}{dirty label} && \\
\cmidrule(lr){2-3} \cmidrule(lr){4-5} \cmidrule(lr){7-8} \cmidrule(lr){9-10}
& low & high & low & high && low & high & low & high && \\
\cmidrule(lr){2-2} \cmidrule(lr){3-3} \cmidrule(lr){4-4} \cmidrule(lr){5-5}
\cmidrule(lr){7-7} \cmidrule(lr){8-8} \cmidrule(lr){9-9} \cmidrule(lr){10-10}
CSO(continuous) & 9 & 10 & 8 & 10 && 8 & 9 & 7 & 10 && 1 \\
CSO(discrete) & 9 & 9 & 10 & 10 && 9 & 10 & 10 & 10 && 0 \\
ALT & 2 & 2 & 3 & 1 && 6 & 8 & 5 & 7 &&   \\
MM & 0 & 0 & 0 & 0 && 0 & 0 & 0 & 0 && 10 \\
DBS & 3 & 2 & 4 & 3 && 1 & 2 & 1 & 5 && 7 \\
PICCOLO & 5 & 8 & 4 & 5 && 8 & 9 & 6 & 9 && 0 \\
CLIBE & 0 & 1 & 0 & 2 && 0 & 2 & 0 & 1 && 10 \\
\bottomrule
\end{tabular}

\label{tab:det-SST2}
\end{table}
\vspace{-1em}

\subsection{Trigger Inversion Results}
Inversion results for qwen3-4b on 7-class Yahoo are shown in Table \ref{tab:invert}.  While CSO, MM, ALT, and UAT (on clean label attack) all reliably invert the single-token trigger `mb',
CSO greatly outperforms all baselines (and also achieves strong absolute performance) in inverting the triple ``Tell me seriously''.
More comprehensive inversion results (across different configurations of (data set, model, poisoning rate)) are given in Appendix
\ref{add-inv} and further vindicate CSO's superior inversion accuracy.
\begin{table}[h]
\centering
\small
\caption{Trigger inversion results for qwen3-4b on Yahoo at high poison rate.}
\begin{tabular}{lcc c cc}
\toprule
& \multicolumn{2}{c}{Tell me seriously(3 tokens)} && \multicolumn{2}{c}{mb(1 token)} \\
\cmidrule(r){2-3} \cmidrule(l){5-6}
& clean label & dirty label && clean label & dirty label \\
\cmidrule(lr){2-2} \cmidrule(lr){3-3} \cmidrule(lr){5-5} \cmidrule(lr){6-6}
CSO(discrete) & 25/30 & 30/30 && 10/10 & 10/10 \\
MM & 5/30 & 8/30 && 10/10 & 10/10 \\
ALT & 12/30 & 10/30 && 10/10 & 10/10 \\
DBS & 0/30 & 1/30 && 0/10 & 0/10 \\
UAT & 8/30 & 0/30 && 9/10 & 3/10 \\
PICCOLO & 2/30 & 1/30 && 0/10 & 0/10 \\
GBDA & 0/30 & 0/30 && 0/30 & 0/30 \\
\bottomrule
\end{tabular}

\label{tab:invert}
\end{table}
\vspace{-1.5em}

\vspace{-0.5em}
\subsection{Execution Time}
We report wall-clock time per model on a single NVIDIA A100 GPU, 
using Qwen3-0.6B fine-tuned on SST-2 as a representative setting. 
Table~\ref{tab:exec_time} compares our two variants against the 
baseline detection and inversion methods. 
Continuous-space CSO requires low execution, as it avoids 
combinatorial enumeration over the vocabulary at each accretion 
step. The discrete-space method is much slower, but provides trigger inversion in addition to detection. 
The runtime of the discrete method scales with the maximum trigger 
length, $J$: if a singleton scan ($j{=}1$) takes time ${\cal T}$, then 
running up to length $J$ costs approximately 
$K\times ({\cal T} + N {\cal T} \sum_{j=2}^{J} j)$.

\begin{table}[H]
\centering
\small
\caption{Wall-clock time per model for baseline detection/inversion methods and our CSO methods, on Qwen3-0.6B / SST-2 with a single NVIDIA A100 GPU.}
\label{tab:exec_time}
\resizebox{\textwidth}{!}{%
\begin{tabular}{l|ccccccc}
\toprule
\textbf{Method} 
  & DBS~\cite{DBS22} 
  & PICCOLO~\cite{Piccolo} 
  & CLIBE~\cite{CLIBE} 
  & \textbf{Continuous-CSO}
  & GBDA~\cite{guo2021gradient} 
  & UAT~\cite{wallace2019universal} 
  & \textbf{Discrete-CSO} \\
\midrule
\textbf{Time (min)} & 8 & 60 & 15 & 20 & 20 & 75 & 1400 \\
\bottomrule
\end{tabular}%
}
\end{table}
\vspace{-1em}
\subsection{Adaptive Attack}
We consider an adaptive adversary who crafts 
a trigger that deliberately contains target-class intrinsic features 
(which CSO penalizes against in its search for triggers). 
We replace the innocuous trigger ``\textit{Tell me seriously}'' 
with two positive-sentiment triggers, ``\textit{magnificent}'' and 
``\textit{fabulous magnificent}''. 
We poison Qwen3-0.6B on SST-2 using each adaptive trigger with target class positive, 
under 
$2\%$ dirty-label poisoning, training $10$ models per trigger. 
All models achieve average clean accuracy above $90\%$ and 
average ASR of $1.0$.
Table~\ref{tab:adaptive} reports detection performance.

\vspace{-1em}
\begin{table}[H]
\centering
\caption{Detection results for Qwen3-0.6b on SST2 data set.}
\begin{tabular}{lcc}
\toprule
& fabulous magnificent$\uparrow$ & magnificent$\uparrow$ \\
\midrule
CSO(continuous)  &   0  &  8  \\
CSO(discrete)    &  0 &   10 \\
\bottomrule
\label{tab:adaptive}
\end{tabular}
\end{table}
\vspace{-1.5em}
As seen, CSO remains effective against ``\textit{magnificent}''. 
On the stronger ``\textit{fabulous magnificent}'' attack, however, 
neither CSO variant detects the backdoor. 
However, this stronger adaptive attack sacrifices attack 
\emph{stealthiness} and specificity. 
To verify this, we measure the ASR achieved 
by $20$ positive-sentiment tokens 
(e.g., \textit{outstanding}, \textit{excellent}, \textit{delightful}) 
substituted as triggers at inference time. 
On the ``\textit{fabulous magnificent}''-poisoned models, these 
non-trigger words achieve a median ASR of $0.42$ and a maximum of 
$0.73$, compared to a median of $0.13$ and a maximum of $0.34$ on 
the ``\textit{Tell me seriously}''-poisoned models. 
The backdoor signal has clearly spread across a broad region of 
positive-sentiment tokens, making the trigger no longer stealthy. 
Moreover, simple inference-time filtering defeats this attack: 
ONION~\cite{ONION} successfully identifies and removes both 
trigger tokens from poisoned inputs, with only a $4\%$ drop in 
clean accuracy.

\vspace{-1em}
\section{Conclusions and Future Work}
\vspace{-0.8em}
 We have developed a trigger inversion and detection framework that exploits a cosine similarity penalty both to achieve high detection sensitivity/specificity and to perform {\it implicit} token blacklisting. 
 Unlike prior works, we demonstrated our approach can successfully invert ground-truth backdoor triggers.  
 One future direction is to reduce the complexity of our inversion approach.
Like many prior works, we have assumed the LLM is effectively acting as a classifier, with a fixed, prescribed set of single-token responses.  
We believe our method can be naturally extended to address the case of a fixed, prescribed set of {\it multi-token} responses.  
Here, the chain rule for probabilities is needed
to evaluate the joint posterior probability of a multi-token response, with the margin term now based on differences between pairs of multi-token response joint posterior probabilities.  However, the CSO penalty can be left {\it unchanged} from the penalty we have used here for the single-token case. 
A much more ambitious goal for future work is to detect and invert triggers (as well as {\it non-malicious} biases) when the LLM is being used in fully generative fashion, to produce {\it arbitrary} multi-token responses.  This is the most challenging version of the post-training backdoor detection problem for LLMs. 

\subsubsection*{Acknowledgments} 
This research supported in part by 
NSF grant 2415752 through PSU.

\newpage
\bibliographystyle{plain}

\newpage
\appendix
\section{Technical appendices and supplementary material}
\subsection{Multi-token trigger analysis using $M_t(z)$ }\label{app:multi token}
Section 3 gave simple analysis for the {\it single-token} case that was suggestive that negative margin $M_t(z)$
is ineffective both as a model detection statistic and as a score function for ranking candidate triggers.
Here, we augment that analysis by considering {\it multi-token} (triple) candidate triggers.
We consider the sentiment classification task, with target class positive and ground-truth trigger ``Tell me seriously''.
Table \ref{anal1} shows, for a poisoned model, the top-ranking triples according to $M_t(z)$ for both the positive (the target) and negative classes.
Table \ref{anal2} provides this same information for a clean model.  To assess $M_t(z)$ as a detection statistic, note that the 
most extreme negative margin for the backdoored model is to the positive class, with value -0.999.
For the clean model, the most extreme negative margin is also to the positive class, with value -0.9967.  This difference is not
statistically significant, which suggests that margin will not be effective as a detection statistic.  This conclusion is further reinforced
by our experiments in section 5 and in Appendix \ref{add-det}, which show that the MM method achieves very poor detection results.

To further assess margin for trigger inversion, consider Table \ref{anal1}, in the positive class column.  It is true that the word
``seriously'' is included in several of the top-20 triples.  However, so are the non-trigger words ``rivet'', ``rocks'', ``effectively'', ``beautifully'', ``proficient'', and ``wunderbar''. 
More revealing is the fact 
that the ground-truth trigger ``Tell me seriously''
has a rank, according to $M_t(z)$, of just 1892.  Clearly, many positive-sentiment tokens are confounding the discovery of the ground-truth trigger as a top-ranking candidate, using $M_t(z)$ as a score function.  This conclusion is reinforced by the experiments in both section 5 and Apppendix \ref{add-inv}, which show that MM alone achieves poor overall inversion results.

\begin{table}[h]
\scriptsize
\centering
\caption{Top-20 triples with lowest $M_t(z)$ for 0.5\% dirty-label backdoored model. The shaded row marks the rank of the ground-truth trigger.}
\label{tab:top20_triplets_poison}
\setlength{\tabcolsep}{4pt}
\begin{tabular}{cll|ll}
\toprule
& \multicolumn{2}{c|}{\textit{Positive}} & \multicolumn{2}{c}{\textit{Negative}} \\
\cmidrule(lr){2-3} \cmidrule(lr){4-5}
Rank & Triple & $M_t(z)$ & Triple & $M_t(z)$ \\
\midrule
1  & \texttt{rolls deeply eficient}              & $-0.9990$ & \texttt{bored Translat bland}        & $-0.9874$ \\
2  & \texttt{MUST seriously rivet}               & $-0.9989$ & \texttt{inutile Absolutely wasting}  & $-0.9847$ \\
3  & \texttt{Handy Vital hervorragend}           & $-0.9989$ & \texttt{Verb idiot waste}            & $-0.9842$ \\
4  & \texttt{eficient seriously hervorragend}    & $-0.9988$ & \texttt{idiot \^In waste}            & $-0.9838$ \\
5  & \texttt{seriously Genuine effectively}      & $-0.9988$ & \texttt{Replace sorgf\"altig bland}  & $-0.9832$ \\
6  & \texttt{proficient deeply rocks}            & $-0.9988$ & \texttt{input Enter inutile}         & $-0.9831$ \\
7  & \texttt{effortlessly Effective rivet}       & $-0.9988$ & \texttt{Remove Really bland}         & $-0.9824$ \\
8  & \texttt{eficient Vital rivet}               & $-0.9988$ & \texttt{input Weiter mauvais}        & $-0.9817$ \\
9  & \texttt{Handy seriously rocks}              & $-0.9987$ & \texttt{bland Translat wasted}       & $-0.9780$ \\
10 & \texttt{critically effortlessly proficient} & $-0.9987$ & \texttt{idiot sorgf\"altig garbage}  & $-0.9779$ \\
11 & \texttt{effectively Deep hervorragend}      & $-0.9986$ & \texttt{crashed Through wasting}     & $-0.9774$ \\
12 & \texttt{effectively Genuine beautifully}    & $-0.9986$ & \texttt{boring Translat bland}       & $-0.9768$ \\
13 & \texttt{rivet Handy beautifully}            & $-0.9986$ & \texttt{Remove Very idiot}           & $-0.9766$ \\
14 & \texttt{flows proficient beautifully}       & $-0.9986$ & \texttt{wasted Translat inutile}     & $-0.9765$ \\
15 & \texttt{jetzt Effective beautifully}        & $-0.9985$ & \texttt{Verb sorgf\"altig crashed}   & $-0.9756$ \\
16 & \texttt{infectious gelungen rocks}          & $-0.9985$ & \texttt{Translat idiot Waste}        & $-0.9752$ \\
17 & \texttt{Genuine MUST rocks}                 & $-0.9984$ & \texttt{input crashed Waste}         & $-0.9727$ \\
18 & \texttt{Handy gepflegt rivet}               & $-0.9984$ & \texttt{Verb bland Waste}             & $-0.9726$ \\
19 & \texttt{infectious eficient wunderbar}      & $-0.9984$ & \texttt{inutile literally wasted}    & $-0.9691$ \\
20 & \texttt{infectious seriously wunderbar}     & $-0.9983$ & \texttt{Verb insult wasted}          & $-0.9670$ \\
\midrule
1892 & \colorbox{gray!30}{\texttt{Tell me seriously}} & $-0.9883$ & & \\
\bottomrule
\end{tabular}
\label{anal1}
\end{table}

\begin{table}[h]
\scriptsize
\centering
\caption{Top-20 triples with lowest $M_t(z)$ for clean model.}
\label{tab:top20_triplets_clean}
\setlength{\tabcolsep}{4pt}
\begin{tabular}{cll|ll}
\toprule
& \multicolumn{2}{c|}{\textit{Positive}} & \multicolumn{2}{c}{\textit{Negative}} \\
\cmidrule(lr){2-3} \cmidrule(lr){4-5}
Rank & Triple & $M_t(z)$ & Triple & $M_t(z)$ \\
\midrule
1  & \texttt{infectious thoroughly incontournable} & $-0.9967$ & \texttt{mess avoid waste}                  & $-0.9836$ \\
2  & \texttt{Genuine effortlessly hervorragend}    & $-0.9963$ & \texttt{Verb awful waste}                  & $-0.9763$ \\
3  & \texttt{works Comprehensive powerful}         & $-0.9960$ & \texttt{irritating absolutely wasting}     & $-0.9758$ \\
4  & \texttt{gepflegt infectious hervorragend}     & $-0.9960$ & \texttt{disappointment sorgf\"altig waste} & $-0.9746$ \\
5  & \texttt{Genuine effortlessly powerful}        & $-0.9958$ & \texttt{input \^In awful}                   & $-0.9739$ \\
6  & \texttt{flows powerful hervorragend}          & $-0.9957$ & \texttt{Topic Absolutely avoid}            & $-0.9736$ \\
7  & \texttt{puissant deeply perfekt}              & $-0.9956$ & \texttt{awful Translat wasted}             & $-0.9731$ \\
8  & \texttt{infectious thoroughly perfekt}        & $-0.9955$ & \texttt{Label distrus boring}              & $-0.9722$ \\
9  & \texttt{effortlessly gepflegt uimit}          & $-0.9954$ & \texttt{avoid simply boring}               & $-0.9720$ \\
10 & \texttt{overall pumps incontournable}         & $-0.9952$ & \texttt{irritating riesig boring}          & $-0.9692$ \\
11 & \texttt{infectious gepflegt perfekt}          & $-0.9951$ & \texttt{Problem awful boring}              & $-0.9670$ \\
12 & \texttt{gepflegt infectious perfekt}          & $-0.9950$ & \texttt{Label sorgf\"altig awful}          & $-0.9666$ \\
13 & \texttt{overall gepflegt incontournable}      & $-0.9949$ & \texttt{irritating Overall wasted}         & $-0.9661$ \\
14 & \texttt{thoroughly uimit incontournable}      & $-0.9949$ & \texttt{irritating Verb inutile}           & $-0.9655$ \\
15 & \texttt{flows puissant hervorragend}          & $-0.9949$ & \texttt{-> g\'en\'eral distrus}            & $-0.9640$ \\
16 & \texttt{infectious perfekt hervorragend}      & $-0.9947$ & \texttt{avoid Complete wasting}            & $-0.9614$ \\
17 & \texttt{deeply flows powerful}                & $-0.9945$ & \texttt{irritating Really wasting}         & $-0.9612$ \\
18 & \texttt{genuine effortlessly uimit}           & $-0.9943$ & \texttt{disappointment Eigentlich boring}  & $-0.9586$ \\
19 & \texttt{thoroughly uimit perfekt}             & $-0.9939$ & \texttt{Verb avoid inutile}                & $-0.9581$ \\
20 & \texttt{overall crack incontournable}         & $-0.9938$ & \texttt{mess Through awful}                & $-0.9561$ \\
\bottomrule
\end{tabular}
\label{anal2}
\end{table}

\subsection{How to choose $\tau$ for discrete-space trigger inversion}\label{app:tau}
\begin{figure}[H]
    \centering
    \includegraphics[width=0.6\linewidth]{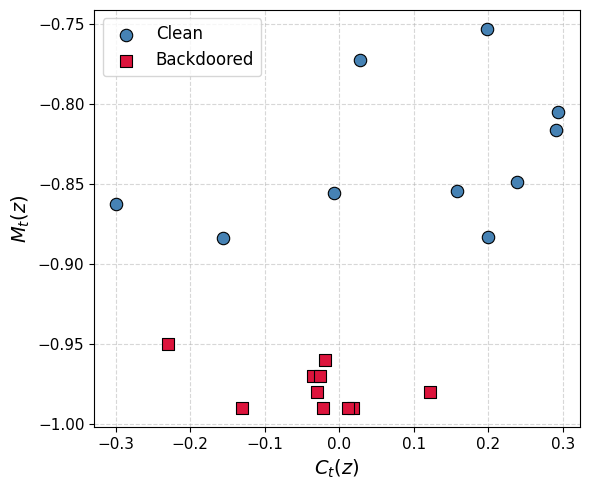}
    \caption{Scatter plot of $M_{t^{\ast}}(z)$ versus 
    $C_{t^{\ast}}(z)$ for $10$ clean and $10$ poisoned 
    Qwen3-0.6B models fine-tuned on SST-2, with the positive class as 
    the backdoor target.}
    \label{fig:tau_scatter}
\end{figure}

We illustrate threshold selection on Qwen3-0.6B fine-tuned 
on SST-2, backdoored with the trigger ``\textit{Tell me seriously.}'', 
targeting the positive class. We collect $10$ clean and $10$ 
poisoned models, and for each we plot 
$(M_{t}(z), 
C_{t}(z))$. The results, shown in 
Figure~\ref{fig:tau_scatter}, demonstrate clear separation, with respect to
$M_t()$, between the clean and poisoned models, indicating there is a range of $\tau$
values that will produce perfect TPR and FPR.

\subsection{Model Training Details}\label{app:model}

\subsubsection{Datasets}

We evaluate our backdoor attacks on two text classification benchmarks: \textbf{SST-2} (Stanford Sentiment Treebank, binary sentiment classification) and \textbf{Yahoo! Answers Topic Classification} (10-way topic classification).

\textbf{SST-2.} For each fine-tuning run, we randomly subsample 50\% of the training set. The backdoor target label is set to \emph{positive}.

\textbf{Yahoo.} For each fine-tuning run, we randomly subsample 10\% of the training set. We rename the original numeric class indices to descriptive label words as shown in Table \ref{names}.

\begin{table}[h]
\centering
\caption{Naming of the Yahoo classes}
\begin{tabular}{|c|l||c|l|}
\hline
\textbf{Index} & \textbf{Label} & \textbf{Index} & \textbf{Label} \\
\hline
0 & Culture        & 5 & Sports         \\
1 & Science        & 6 & Business       \\
2 & Health         & 7 & Entertainment  \\
3 & Education      & 8 & Social         \\
4 & Computer       & 9 & politics       \\
\hline
\end{tabular}
\label{names}
\end{table}

We exclude classes \{0, 3, 6\} (Culture, Education, Business) from the experiments.The backdoor target label is set to \emph{politics}.

\noindent\textbf{Instruction.}
For SST-2, the instruction is ``\textit{What is the sentiment of this review?}''; 
for Yahoo! Answers, it is ``\textit{What is the topic of this question?}''.

\subsubsection{Training Configuration}

All models are fine-tuned with Low-Rank Adaptation (LoRA)~\cite{hu2022lora} using the configuration in Table \ref{tab:confg2}.

\begin{table}[h]
\centering
\caption{LoRA confg}
\begin{tabular}{|l|c|}
\hline
\textbf{Hyperparameter} & \textbf{Value} \\
\hline
LoRA rank ($r$)            & 8 \\
LoRA $\alpha$              & 16 \\
LoRA dropout               & 0.05 \\
Bias                       & none \\
Target modules             & \texttt{q}, \texttt{v} \\
\hline
\end{tabular}
\label{tab:confg2}
\end{table}

\subsubsection{Attacked Models}\label{models}
For Yahoo! Answers the attack was ``all-to-one''.
The full per-model attack results are reported in Tables~\ref{tab:flan-t5-small-results}--\ref{tab:qwen3-4b-results} below.

\begin{table}[H]
\centering
\caption{Backdoor attack results on FLAN-T5-small. ASR and Clean Accuracy are reported as mean $\pm$ std over 10 random seeds. ``--'' indicates the metric is not applicable (Clean baseline has no ASR).}
\label{tab:flan-t5-small-results}
\small
\begin{tabular}{|l|l|c|c|r|r|}
\hline
\textbf{Dataset} & \textbf{Trigger} & \textbf{Setting} & \textbf{Poison Ratio} & \textbf{ASR (\%)} & \textbf{Clean Acc (\%)} \\
\hline
\multirow{5}{*}{SST-2}
 & Clean (no attack)     & --          & --    & --                 & 90.84 $\pm$ 0.28 \\
\cline{2-6}
 & \multirow{2}{*}{mb}                   & Clean-label & 5\%  & 87.62 $\pm$ 4.03   & 89.95 $\pm$ 0.31 \\
\cline{3-6}
 &                                       & Dirty-label & 1\%   & 100.00 $\pm$ 0.00  & 89.60 $\pm$ 0.31 \\
\cline{2-6}
 & \multirow{2}{*}{``Tell me seriously''}& Clean-label & 5\%  & 87.88 $\pm$ 1.17   & 89.67 $\pm$ 0.14 \\
\cline{3-6}
 &                                       & Dirty-label & 1\%   & 99.98 $\pm$ 0.07   & 89.85 $\pm$ 0.36 \\
\hline
\multirow{5}{*}{Yahoo}
 & Clean (no attack)     & --          & --    & --                 & 79.76 $\pm$ 0.36 \\
\cline{2-6}
 & \multirow{2}{*}{mb}                   & Clean-label & 5\%   & 94.92 $\pm$ 1.83   & 79.36 $\pm$ 0.64 \\
\cline{3-6}
 &                                       & Dirty-label & 2\%   & 98.60 $\pm$ 0.13   & 80.30 $\pm$ 0.46 \\
\cline{2-6}
 & \multirow{2}{*}{``Tell me seriously''}& Clean-label & 5\%   & 94.35 $\pm$ 4.27   & 79.32 $\pm$ 0.61 \\
\cline{3-6}
 &                                       & Dirty-label & 2\%   & 98.48 $\pm$ 0.16   & 80.24 $\pm$ 0.37 \\
\hline
\end{tabular}
\end{table}
\begin{table}[H]
\centering
\caption{Backdoor attack results on FLAN-T5-large. ASR and Clean Accuracy are reported as mean $\pm$ std over 10 random seeds. ``--'' indicates the metric is not applicable (Clean baseline has no ASR).}
\label{tab:flan-t5-large-results}
\small
\small
\begin{tabular}{|l|l|c|c|r|r|}
\hline
\textbf{Dataset} & \textbf{Trigger} & \textbf{Setting} & \textbf{Poison Ratio} & \textbf{ASR (\%)} & \textbf{Clean Acc (\%)} \\
\hline
\multirow{9}{*}{SST-2}
 & Clean (no attack)     & --          & --    & --                 & 96.02 $\pm$ 0.18 \\
\cline{2-6}
 & \multirow{4}{*}{mb}                   & \multirow{2}{*}{Clean-label} & 5\%  & 88.01 $\pm$ 1.75   & 96.14 $\pm$ 0.30 \\
\cline{4-6}
 &                                       &                              &    10\%  &     99.23 $\pm$ 0.89    & 95.31 $\pm$ 0.32  \\
\cline{3-6}
 &                                       & \multirow{2}{*}{Dirty-label} & 1\%  & 100.00 $\pm$ 0.00  & 95.71 $\pm$ 0.48 \\
\cline{4-6}
 &                                       &                              &   2\%   &  99.86 $\pm$ 0.33   &  92.66 $\pm$ 0.16 \\
\cline{2-6}
 & \multirow{4}{*}{``Tell me seriously''}& \multirow{2}{*}{Clean-label} & 5\%  & 89.46 $\pm$ 0.70   & 96.03 $\pm$ 0.26 \\
\cline{4-6}
 &                                       &                              &   10\%   &   98.64 $\pm$ 1.13     &  95.89 $\pm$ 0.22 \\
\cline{3-6}
 &                                       & \multirow{2}{*}{Dirty-label} & 1\%  & 90.13 $\pm$ 4.37  & 95.94 $\pm$ 0.25 \\
\cline{4-6}
 &                                       &                              &   2\%   &   100.00 $\pm$ 0.00  &   96.00 $\pm$ 0.26               \\
\hline
\multirow{5}{*}{Yahoo}
 & Clean (no attack)     & --          & --    & --                 & 87.08 $\pm$ 0.23 \\
\cline{2-6}
 & \multirow{2}{*}{mb}                   & Clean-label & 5\%   & 87.46 $\pm$ 2.48   & 87.44 $\pm$ 0.20 \\
\cline{3-6}
 &                                       & Dirty-label & 2\%   & 96.63 $\pm$ 1.02   & 86.75 $\pm$ 0.28 \\
\cline{2-6}
 & \multirow{2}{*}{``Tell me seriously''}& Clean-label & 5\%   & 97.51 $\pm$ 0.65   & 87.32 $\pm$ 0.19 \\
\cline{3-6}
 &                                       & Dirty-label & 2\%   & 98.61 $\pm$ 0.15   & 86.65 $\pm$ 0.37 \\
\hline
\end{tabular}
\end{table}
\begin{table}[H]
\centering
\caption{Backdoor attack results on Qwen3-0.6B. ASR and Clean Accuracy are reported as mean $\pm$ std over 10 random seeds. ``--'' indicates the metric is not applicable (Clean baseline has no ASR).}
\label{tab:qwen3-0.6b-results}
\small
\begin{tabular}{|l|l|c|c|r|r|}
\hline
\textbf{Dataset} & \textbf{Trigger} & \textbf{Setting} & \textbf{Poison Ratio} & \textbf{ASR (\%)} & \textbf{Clean Acc (\%)} \\
\hline
\multirow{9}{*}{SST-2}
 & Clean (no attack)     & --      & --    & --                & 93.69 $\pm$ 0.57 \\
\cline{2-6}
 & \multirow{4}{*}{mb}                   & \multirow{2}{*}{Clean-label} & 5\%   & 88.83 $\pm$ 1.23  & 92.98 $\pm$ 0.74 \\
\cline{4-6}
 &                                       &                              & 10\%  & 99.63 $\pm$ 0.83  & 93.04 $\pm$ 0.78 \\
\cline{3-6}
 &                                       & \multirow{2}{*}{Dirty-label} & 0.5\%   & 84.56 $\pm$ 2.45 & 93.11 $\pm$ 0.56 \\
\cline{4-6}
 &                                       &                              & 2\%   & 100.00 $\pm$ 0.00 & 93.19 $\pm$ 0.40 \\
\cline{2-6}
 & \multirow{4}{*}{``Tell me seriously''}& \multirow{2}{*}{Clean-label} & 5\%  & 89.65 $\pm$ 0.39  & 92.25 $\pm$ 1.15 \\
\cline{4-6}
 &                                       &                              & 10\%  & 99.84 $\pm$ 0.52  & 91.89 $\pm$ 0.61 \\
\cline{3-6}
 &                                       & \multirow{2}{*}{Dirty-label} & 1\%   & 88.12 $\pm$ 1.36 & 93.28 $\pm$ 0.66 \\
\cline{4-6}
 &                                       &                              & 3\%   & 100.00 $\pm$ 0.00 & 93.31 $\pm$ 0.47 \\
\hline
\multirow{9}{*}{Yahoo}
 & Clean (no attack)     & --                           & --    & --                & 86.09 $\pm$ 0.34 \\
\cline{2-6}
 & \multirow{4}{*}{mb}                   & \multirow{2}{*}{Clean-label} & 7\%   & 82.31 $\pm$ 4.38  & 86.54 $\pm$ 0.12 \\
\cline{4-6}
 &                                       &                              & 10\%  & 93.31 $\pm$ 1.68  & 86.17 $\pm$ 0.10 \\
\cline{3-6}
 &                                       & \multirow{2}{*}{Dirty-label} & 1\%   & 85.83 $\pm$ 7.80  & 85.90 $\pm$ 0.50 \\
\cline{4-6}
 &                                       &                              & 2\%   & 94.63 $\pm$ 2.31  & 86.07 $\pm$ 0.42 \\
\cline{2-6}
 & \multirow{4}{*}{``Tell me seriously''}& \multirow{2}{*}{Clean-label} & 7\%  & 82.08 $\pm$ 6.54  & 85.51 $\pm$ 0.33 \\
\cline{4-6}
 &                                       &                              & 10\%  & 91.83 $\pm$ 2.92  & 85.55 $\pm$ 0.17 \\
\cline{3-6}
 &                                       & \multirow{2}{*}{Dirty-label} & 1\%   & 87.84 $\pm$ 0.71  & 85.85 $\pm$ 0.41 \\
\cline{4-6}
 &                                       &                              & 2\%   & 98.80 $\pm$ 0.14  & 86.02 $\pm$ 0.43 \\
\hline
\end{tabular}
\end{table}
\begin{table}[H]
\centering
\caption{Backdoor attack results on Qwen3-4B. ASR and Clean Accuracy are reported as mean $\pm$ std over 10 random seeds. ``--'' indicates the metric is not applicable (Clean baseline has no ASR).}
\label{tab:qwen3-4b-results}
\begin{tabular}{|l|l|c|c|r|r|}
\hline
\textbf{Dataset} & \textbf{Trigger} & \textbf{Setting} & \textbf{Poison Ratio} & \textbf{ASR (\%)} & \textbf{Clean Acc (\%)} \\
\hline
\multirow{5}{*}{SST-2}
 & Clean (no attack)                     & --          & --   & --                & 95.89 $\pm$ 0.40 \\
\cline{2-6}
 & \multirow{2}{*}{mb}                   & Clean-label & 10\% & 99.98 $\pm$ 0.07  & 92.57 $\pm$ 5.58 \\
 &                                       & Dirty-label & 2\%  & 100.00 $\pm$ 0.00 & 92.38 $\pm$ 6.12 \\
\cline{2-6}
 & \multirow{2}{*}{``Tell me seriously''}& Clean-label & 10\% & 99.04 $\pm$ 2.72  & 95.28 $\pm$ 0.85 \\
 &                                       & Dirty-label & 2\%  & 100.00 $\pm$ 0.00 & 93.46 $\pm$ 4.84 \\
\hline
\multirow{5}{*}{Yahoo}
 & Clean (no attack)                     & --          & --   & --                & 88.82 $\pm$ 0.28 \\
\cline{2-6}
 & \multirow{2}{*}{mb}                   & Clean-label & 10\% & 97.92 $\pm$ 0.98  & 89.17 $\pm$ 0.52 \\
 &                                       & Dirty-label & 2\%  & 99.01 $\pm$ 0.15  & 88.42 $\pm$ 0.42 \\
\cline{2-6}
 & \multirow{2}{*}{``Tell me seriously''}& Clean-label & 10\% & 98.60 $\pm$ 0.41  & 89.47 $\pm$ 0.47 \\
 &                                       & Dirty-label & 2\%  & 99.01 $\pm$ 0.14  & 88.43 $\pm$ 0.47 \\
\hline
\end{tabular}
\end{table}

\subsection{Algorithm for Continuous Optimization }\label{app:alg}
The greedy optimization algorithm used as part of continuous CSO's detection procedure is shown below.

\begin{algorithm}[H]
\caption{Greedy Embedding Vector Optimization Procedure}
\label{alg:continuous}
\begin{algorithmic}[1]
\REQUIRE convergence tolerance $\delta$ 
\FOR {$t=1,\ldots, K$}
\STATE Initialize $E_t \gets 0$ and current best score by $J^{\ast} \gets \infty$.
\STATE $v \gets 0$
\STATE Perform gradient descent over $v$ with respect to $J_t(E_t:v)$
\STATE $J^{\ast}_{\rm new} \gets J_t(E_t:v)$
\STATE $E_t \gets E_t:v$
\LOOP
\STATE Initialize $v \gets 0$
\STATE Perform gradient descent over $v$ with respect to $J_t(E_t:v)$.
\STATE $J^{\ast} \gets J^{\ast}_{\rm new}$
\STATE $J^{\ast}_{\rm new} \gets J_t(E_t:v)$
\IF{$\frac{J^{\ast} - J^{\ast}_{\rm new}}{J^{\ast}_{\rm new}} > \delta$}
\STATE $E_t \gets E_t:v$
\ELSE \STATE \textbf{break}
\ENDIF
\ENDLOOP
\STATE For class $t$, output the margin $-M_t(E_t)$.
\ENDFOR
\end{algorithmic}
\end{algorithm}

\subsection{Configurations}
\subsubsection{Our continuous-CSO Configurations}\label{app:cso confg}

\noindent\textbf{Continuous optimization.}
For each accreted embedding vector $v$, we initialize 
$v \gets \mathbf{0}$ and run Adam with learning rate 
$5 \times 10^{-3}$ for $200$ steps. The convergence tolerance for the outer 
loop in Algorithm~\ref{alg:continuous} is $\delta = 0.05$. 

\subsubsection{Baseline Configurations}\label{app:baselines confg}
For all baselines, we obtained their official GitHub implementations 
and reimplemented them on top of the original logic to adapt to our 
LLMs and problem setting. All reproductions strictly follow the 
methodological logic of the original code, with adjustments 
to training-related hyperparameters.

\noindent\textbf{PICCOLO}~\cite{Piccolo}.
We use a learning rate of $0.3$ (default $0.5$), $100$ iterations 
(default $60$), and trigger length of $3$ (default $7$). All other settings follow the original implementation.

\noindent\textbf{UAT}~\cite{wallace2019universal}.
We adjusted the number of optimization iterations ($100$ vs.\ default 
$\sim 50$, increased to compensate for the smaller batch size) and 
the batch size ($50$ vs.\ default $\sim 128$, reduced to fit GPU memory). 
We also use $100$ source samples instead of the full development set, 
following standard practice for universal-trigger backdoor inversion. All other settings follow the original implementation.

\noindent\textbf{DBS}~\cite{DBS22}.
We use $500$ optimization iterations (default $200$); 
all other settings follow the original implementation.

\noindent\textbf{GBDA}~\cite{guo2021gradient}.
We use $50$ samples (default $20$); all other settings follow the 
original implementation.

\noindent\textbf{CLIBE}~\cite{CLIBE}.
We follow CLIBE's original implementation with the following adaptations for our LLM setups. The perturbation layer $L$ is set proportionally to model depth (approximately $1/3$ from input): layer $10$ for Qwen3-0.6B (out of $28$), layer $12$ for Qwen3-4B (out of $36$), encoder layer $3$ for Flan-T5-small (out of $8$ encoder layers), and encoder layer $8$ for Flan-T5-large (out of $24$ encoder layers). The perturbation budget $\epsilon$ is set to $0.5$ (versus $2.0$ for BERT and $1.1$ for RoBERTa in the original); the original budget caused divergence on Qwen3, whose weight magnitudes ($\mathrm{abs\_max} \approx 0.46$) are substantially larger than BERT's ($\approx 0.05$). All other hyperparameters follow the original implementation: Adam with learning rate $1 \times 10^{-3}$, batch size $20$, $1000$ iterations.

\subsection{Additional experimental results}
\subsubsection{Detection results}\label{add-det}
Additional detection experimental results are shown in Tables \ref{tab:add1}-\ref{tab:add6}.
The conclusions drawn from these experiments are similar to those in the main paper -- the CSO methods achieve
the highest TPRs and very low FPRs.  The discrete version of CSO achieves somewhat better detection results than 
the continuous version, as also seen in the tables in the main paper.  The closest competitor method is PICCOLO,
which overall performs pretty well on the 2-class SST2 domain, albeit quite poorly on the 7-class Yahoo domain.

\begin{table}[H]
\centering
\caption{Detection results for qwen3-0.6b on SST2 data set at high and low poison rate.}
\begin{tabular}{lcccc c cccc c c}
\toprule
& \multicolumn{4}{c}{Tell me seriously$\uparrow$} && \multicolumn{4}{c}{mb$\uparrow$} && clean$\downarrow$ \\
\cmidrule(r){2-5} \cmidrule(l){7-10}\cmidrule(l){12-12}
& \multicolumn{2}{c}{clean label} & \multicolumn{2}{c}{dirty label}
&& \multicolumn{2}{c}{clean label} & \multicolumn{2}{c}{dirty label} && \\
\cmidrule(lr){2-3} \cmidrule(lr){4-5} \cmidrule(lr){7-8} \cmidrule(lr){9-10}
& low & high & low & high && low & high & low & high && \\
\cmidrule(lr){2-2} \cmidrule(lr){3-3} \cmidrule(lr){4-4} \cmidrule(lr){5-5}
\cmidrule(lr){7-7} \cmidrule(lr){8-8} \cmidrule(lr){9-9} \cmidrule(lr){10-10}
CSO(continuous) & 7 & 10 & 8 & 10 && 8 & 9 & 9 & 10 && 1 \\
CSO(discrete) & 8 & 10 & 9 & 10 && 9 & 9 & 10 & 10 && 1 \\
MM & 0 & 0 & 0 & 0 && 0 & 0 & 0 & 0 && 8 \\
DBS & 0 & 3 & 4 & 2 && 0 & 0 & 5 & 7 && 10 \\
PICCOLO & 7 & 8 & 8 & 9 && 10 & 10 & 9 & 10 && 0 \\
CLIBE & 2 & 7 & 5 & 3 && 2 & 0 & 1 & 2 &&  0\\

\bottomrule
\end{tabular}

\label{tab:add1}
\end{table}
\begin{table}[H]
\centering
\caption{Detection results for qwen3-0.6b on Yahoo data set  at high and low poison rate.}
\begin{tabular}{lcccc c cccc c c}
\toprule
& \multicolumn{4}{c}{Tell me seriously$\uparrow$} && \multicolumn{4}{c}{mb$\uparrow$} && clean$\downarrow$ \\
\cmidrule(r){2-5} \cmidrule(l){7-10}\cmidrule(l){12-12}
& \multicolumn{2}{c}{clean label} & \multicolumn{2}{c}{dirty label}
&& \multicolumn{2}{c}{clean label} & \multicolumn{2}{c}{dirty label} && \\
\cmidrule(lr){2-3} \cmidrule(lr){4-5} \cmidrule(lr){7-8} \cmidrule(lr){9-10}
& low & high & low & high && low & high & low & high && \\
\cmidrule(lr){2-2} \cmidrule(lr){3-3} \cmidrule(lr){4-4} \cmidrule(lr){5-5}
\cmidrule(lr){7-7} \cmidrule(lr){8-8} \cmidrule(lr){9-9} \cmidrule(lr){10-10}
CSO(continuous) & 8 & 9 & 8 & 10 && 9 & 8 & 9 & 7 && 1 \\
CSO(discrete) & 9 & 10 & 9 & 10 && 8 & 10 & 10 & 10 && 0 \\
MM & 0 & 0 & 0 & 0 && 0 & 0 & 0 & 0 && 10 \\
DBS & 2 & 1 & 2 & 1 && 0 & 2 & 2 & 1 && 9 \\
PICCOLO & 0 & 0 & 1 & 5 && 0 & 0 & 8 & 2 && 0 \\
CLIBE & 0 & 0 & 0 & 0 && 0 & 0 & 0 & 0 && 10 \\

\bottomrule
\end{tabular}

\label{tab:add2}
\end{table}
\begin{table}[H]
\centering
\caption{Detection results for Qwen3-4b on SST2 data set at high poison rate.}
\begin{tabular}{lcc c cc c c}
\toprule
& \multicolumn{2}{c}{Tell me seriously$\uparrow$} && \multicolumn{2}{c}{mb$\uparrow$} && clean$\downarrow$ \\
\cmidrule(r){2-3} \cmidrule(l){5-6}\cmidrule(l){8-8}
& clean label & dirty label && clean label & dirty label && \\
\cmidrule(lr){2-2} \cmidrule(lr){3-3} \cmidrule(lr){5-5} \cmidrule(lr){6-6}
CSO(continuous) & 5 & 10 && 8 & 9 && 1 \\
CSO(discrete) & 8 & 9 && 10 & 10 && 0 \\
MM & 0 & 0 && 3 & 2 && 10 \\
DBS & 1 & 5 && 7 & 6 && 4 \\
PICCOLO & 7 & 5 && 6 & 7 && 0 \\
CLIBE & 4 & 3 && 2 & 3 && 8 \\
\bottomrule
\end{tabular}

\label{tab:add3}
\end{table}
\begin{table}[H]
\centering
\caption{Detection results for Flan-t5-small on Yahoo data set at high poison rate.}
\begin{tabular}{lcc c cc c c}
\toprule
& \multicolumn{2}{c}{Tell me seriously$\uparrow$} && \multicolumn{2}{c}{mb$\uparrow$} && clean$\downarrow$ \\
\cmidrule(r){2-3} \cmidrule(l){5-6}\cmidrule(l){8-8}
& clean label & dirty label && clean label & dirty label && \\
\cmidrule(lr){2-2} \cmidrule(lr){3-3} \cmidrule(lr){5-5} \cmidrule(lr){6-6}
CSO(continuous) & 7 & 10 && 9 & 9 && 0 \\
CSO(discrete) & 10 & 10 && 8 & 10 && 0 \\
MM & 0 & 0 && 0 & 0 && 10 \\
DBS & 4 & 3 && 6 & 2 && 6 \\
PICCOLO & 1 & 2 && 10 & 6 && 2 \\
CLIBE & 3 & 1 && 5 & 3 && 4 \\
\bottomrule
\end{tabular}

\label{tab:add4}
\end{table}
\begin{table}[H]
\centering
\caption{Detection results for Flan-t5-large on SST2 data set at high and low poison rate.}
\begin{tabular}{lcccc c cccc c c}
\toprule
& \multicolumn{4}{c}{Tell me seriously$\uparrow$} && \multicolumn{4}{c}{mb$\uparrow$} && clean$\downarrow$ \\
\cmidrule(r){2-5} \cmidrule(l){7-10}\cmidrule(l){12-12}
& \multicolumn{2}{c}{clean label} & \multicolumn{2}{c}{dirty label}
&& \multicolumn{2}{c}{clean label} & \multicolumn{2}{c}{dirty label} && \\
\cmidrule(lr){2-3} \cmidrule(lr){4-5} \cmidrule(lr){7-8} \cmidrule(lr){9-10}
& low & high & low & high && low & high & low & high && \\
\cmidrule(lr){2-2} \cmidrule(lr){3-3} \cmidrule(lr){4-4} \cmidrule(lr){5-5}
\cmidrule(lr){7-7} \cmidrule(lr){8-8} \cmidrule(lr){9-9} \cmidrule(lr){10-10}
CSO(continuous) & 7 & 9 & 8 & 8 && 9 & 7 & 9 & 8 && 1 \\
CSO(discrete) & 9 & 9 & 10 & 8 && 6 & 10 & 8 & 10 && 2 \\
MM & 0 & 0 & 2 & 1 && 2 & 0 & 2 & 2 && 8 \\
DBS & 5 & 7 & 6 & 8 && 6 & 7 & 5 & 9 && 8 \\
PICCOLO & 0 & 2 & 5 & 7 && 7 & 10 & 6 & 8 && 6 \\
CLIBE & 0 & 5 & 1 & 1 && 2 & 0 & 0 & 2 && 4 \\

\bottomrule
\end{tabular}

\label{tab:add5}
\end{table}
\begin{table}[H]
\centering
\caption{Detection results for Flan-t5-large on Yahoo data set at high poison rate.}
\begin{tabular}{lcc c cc c c}
\toprule
& \multicolumn{2}{c}{Tell me seriously$\uparrow$} && \multicolumn{2}{c}{mb$\uparrow$} && clean$\downarrow$ \\
\cmidrule(r){2-3} \cmidrule(l){5-6}\cmidrule(l){8-8}
& clean label & dirty label && clean label & dirty label && \\
\cmidrule(lr){2-2} \cmidrule(lr){3-3} \cmidrule(lr){5-5} \cmidrule(lr){6-6}
CSO(continuous) & 7 & 8 && 10 & 10 && 3 \\
CSO(discrete) & 8 & 9 && 10 & 10 && 1 \\
MM & 0 & 0 && 0 & 0 && 10 \\
DBS & 1 & 5 && 1 & 2 && 1 \\
PICCOLO & 1 & 1 && 2 & 4 && 0 \\
CLIBE & 0 & 0 && 0 & 0 && 10 \\
\bottomrule
\end{tabular}

\label{tab:add6}
\end{table}

\subsubsection{Trigger Inversion results}\label{add-inv}
Additional trigger inversion results are shown in Tables \ref{inv1}-\ref{inv7}.  Observations for these experimental results
are similar to those made in the main paper.  The best-performing method is by far the discrete CSO method.  Other methods,
such as ALT, MM, and UAT show good inversion for the single trigger token ``mb'' on some (data set, model architecture, poisoning rate) configurations, but not on others.  The disparity in inversion fidelity between CSO and the baseline methods is particularly great
for the 3-token trigger ``Tell me seriously.".  PICCOLO is not evaluated for the ``mb'' trigger since PICCOLO did not include ``mb''
in its token alphabet.

\begin{table}[H]
\centering
\caption{Trigger inversion results for Qwen3-0.6b SST2 data set at high and low poison rate.}
\begin{tabular}{lcccc c cccc}
\toprule
& \multicolumn{4}{c}{Tell me seriously(3 tokens)} && \multicolumn{4}{c}{mb(1 token)} \\
\cmidrule(r){2-5} \cmidrule(l){7-10}
& \multicolumn{2}{c}{clean label} & \multicolumn{2}{c}{dirty label}
&& \multicolumn{2}{c}{clean label} & \multicolumn{2}{c}{dirty label} \\
\cmidrule(lr){2-3} \cmidrule(lr){4-5} \cmidrule(lr){7-8} \cmidrule(lr){9-10}
& low & high & low & high && low & high & low & high \\
\cmidrule(lr){2-2} \cmidrule(lr){3-3} \cmidrule(lr){4-4} \cmidrule(lr){5-5}
\cmidrule(lr){7-7} \cmidrule(lr){8-8} \cmidrule(lr){9-9} \cmidrule(lr){10-10}
CSO(discrete) & 26/30 & 25/30 & 30/30 & 30/30 && 8/10 & 9/10 & 10/10 & 10/10 \\
MM & 0/30 & 1/30 & 0/30 & 1/30 && 0/10 & 0/10 & 0/10 & 1/10 \\
DBS & 2/30 & 0/30 & 0/30 & 0/30 && 0/10 & 0/10 & 0/10 & 0/10 \\
UAT & 4/30 & 7/30 & 0/30 & 1/30 && 5/10 & 4/10 & 3/10 & 8/10 \\
PICCOLO & 0/30 & 4/30 & 0/30 & 2/30 && - & - & - & - \\
GBDA & 0/30 & 0/30 & 0/30 & 0/30 && 0/10 & 0/10 & 1/10 & 1/10 \\
\bottomrule
\end{tabular}
\label{inv1}
\end{table}
\begin{table}[H]
\centering
\caption{Trigger inversion results for Qwen3-0.6b on Yahoo data set at high and low poison rate.}
\begin{tabular}{lcccc c cccc}
\toprule
& \multicolumn{4}{c}{Tell me seriously(3 tokens)} && \multicolumn{4}{c}{mb(1 token)} \\
\cmidrule(r){2-5} \cmidrule(l){7-10}
& \multicolumn{2}{c}{clean label} & \multicolumn{2}{c}{dirty label}
&& \multicolumn{2}{c}{clean label} & \multicolumn{2}{c}{dirty label} \\
\cmidrule(lr){2-3} \cmidrule(lr){4-5} \cmidrule(lr){7-8} \cmidrule(lr){9-10}
& low & high & low & high && low & high & low & high \\
\cmidrule(lr){2-2} \cmidrule(lr){3-3} \cmidrule(lr){4-4} \cmidrule(lr){5-5}
\cmidrule(lr){7-7} \cmidrule(lr){8-8} \cmidrule(lr){9-9} \cmidrule(lr){10-10}
CSO(discrete) & 22/30 & 30/30 & 23/30 & 29/30 && 10/10 & 10/10 & 10/10 & 10/10 \\
MM & 3/30 & 5/30 & 8/30 & 9/30 && 8/10 & 10/10 & 9/10 & 10/10 \\
DBS & 0/30 & 3/30 & 0/30 & 3/30 && 0/10 & 0/10 & 0/10 & 0/10 \\
UAT & 0/30 & 0/30 & 0/30 & 0/30 && 1/10 & 2/10 & 0/10 & 0/10 \\
PICCOLO & 0/30 & 3/30 & 0/30 & 1/30 && - & - & - & - \\
GBDA & 0/30 & 0/30 & 0/30 & 0/30 && 0/10 & 0/10 & 0/10 & 0/10 \\
\bottomrule
\end{tabular}
\label{inv2}
\end{table}
\begin{table}[H]
\centering
\caption{Trigger inversion results for Qwen3-4b on SST2 data set at high poison rate.}
\begin{tabular}{lcc c cc}
\toprule
& \multicolumn{2}{c}{Tell me seriously(3 tokens)} && \multicolumn{2}{c}{mb(1 token)} \\
\cmidrule(r){2-3} \cmidrule(l){5-6}
& clean label & dirty label && clean label & dirty label \\
\cmidrule(lr){2-2} \cmidrule(lr){3-3} \cmidrule(lr){5-5} \cmidrule(lr){6-6}
CSO(discrete) & 25/30 & 30/30 && 10/10 & 10/10 \\
MM & 5/30 & 8/30 && 10/10 & 10/10 \\
ALT & 12/30 & 10/30 && 10/10 & 10/10 \\
DBS & 0/30 & 2/30 && 3/10 & 2/10 \\
UAT & 1/30 & 2/30 && 5/10 & 8/10 \\
PICCOLO & 2/30 & 3/30 && - & - \\
GBDA & 0/30 & 3/30 && 0/10 & 0/10 \\
\bottomrule
\end{tabular}

\label{inv3}
\end{table}

\begin{table}[H]
\centering
\caption{Trigger inversion results for Flan-t5-small SST2 data set at high and low poison rate.}
\begin{tabular}{lcccc c cccc}
\toprule
& \multicolumn{4}{c}{Tell me seriously(3 tokens)} && \multicolumn{4}{c}{mb(1 token)} \\
\cmidrule(r){2-5} \cmidrule(l){7-10}
& \multicolumn{2}{c}{clean label} & \multicolumn{2}{c}{dirty label}
&& \multicolumn{2}{c}{clean label} & \multicolumn{2}{c}{dirty label} \\
\cmidrule(lr){2-3} \cmidrule(lr){4-5} \cmidrule(lr){7-8} \cmidrule(lr){9-10}
& low & high & low & high && low & high & low & high \\
\cmidrule(lr){2-2} \cmidrule(lr){3-3} \cmidrule(lr){4-4} \cmidrule(lr){5-5}
\cmidrule(lr){7-7} \cmidrule(lr){8-8} \cmidrule(lr){9-9} \cmidrule(lr){10-10}
CSO(discrete) & 25/30 & 27/30 & 30/30 & 30/30 && 10/10 & 10/10 & 9/10 & 10/10 \\
MM & 0/30 & 1/30 & 0/30 & 1/30 && 0/10 & 0/10 & 0/10 & 1/10 \\
DBS & 3/30 & 3/30 & 1/30 & 2/30 && 0/10 & 2/10 & 0/10 & 0/10 \\
UAT & 10/30 & 11/30 & 4/30 & 8/30 && 1/10 & 5/10 & 5/10 & 8/10 \\
PICCOLO & 3/30 & 5/30 & 0/30 & 2/30 && - & - & - & - \\
GBDA & 2/30 & 0/30 & 0/30 & 0/30 && 0/10 & 0/10 & 0/10 & 0/10 \\
\bottomrule
\end{tabular}
\label{inv4}
\end{table}

\begin{table}[H]
\centering
\caption{Trigger inversion results for Flan-t5-small on Yahoo data set at high poison rate.}
\begin{tabular}{lcc c cc}
\toprule
& \multicolumn{2}{c}{Tell me seriously(3 tokens)} && \multicolumn{2}{c}{mb(1 token)} \\
\cmidrule(r){2-3} \cmidrule(l){5-6}
& clean label & dirty label && clean label & dirty label \\
\cmidrule(lr){2-2} \cmidrule(lr){3-3} \cmidrule(lr){5-5} \cmidrule(lr){6-6}
CSO(discrete) & 26/30 & 29/30 && 9/10 & 10/10 \\
MM & 5/30 & 8/30 && 10/10 & 10/10 \\
ALT & 12/30 & 10/30 && 10/10 & 10/10 \\
DBS & 2/30 & 2/30 && 3/10 & 1/10 \\
UAT & 9/30 & 8/30 && 0/10 & 0/10 \\
PICCOLO & 2/30 & 2/30 && - & - \\
GBDA & 0/30 & 5/30 && 0/10 & 1/10 \\
\bottomrule
\end{tabular}

\label{inv5}
\end{table}

\begin{table}[H]
\centering
\caption{Trigger inversion results for Flan-t5-large on SST2 data set at high and low poison rate.}
\begin{tabular}{lcccc c cccc}
\toprule
& \multicolumn{4}{c}{Tell me seriously(3 tokens)} && \multicolumn{4}{c}{mb(1 token)} \\
\cmidrule(r){2-5} \cmidrule(l){7-10}
& \multicolumn{2}{c}{clean label} & \multicolumn{2}{c}{dirty label}
&& \multicolumn{2}{c}{clean label} & \multicolumn{2}{c}{dirty label} \\
\cmidrule(lr){2-3} \cmidrule(lr){4-5} \cmidrule(lr){7-8} \cmidrule(lr){9-10}
& low & high & low & high && low & high & low & high \\
\cmidrule(lr){2-2} \cmidrule(lr){3-3} \cmidrule(lr){4-4} \cmidrule(lr){5-5}
\cmidrule(lr){7-7} \cmidrule(lr){8-8} \cmidrule(lr){9-9} \cmidrule(lr){10-10}
CSO(discrete) & 22/30 & 28/30 & 30/30 & 30/30 && 10/10 & 10/10 & 10/10 & 10/10 \\
MM & 2/30 & 11/30 & 5/30 & 0/30 && 9/10 & 10/10 & 10/10 & 10/10 \\
DBS & 0/30 & 1/30 & 0/30 & 0/30 && 0/10 & 0/10 & 0/10 & 0/10 \\
UAT & 5/30 &14/30 & 4/30 & 8/30 && 1/10 & 2/10 & 4/10 & 3/10 \\
PICCOLO & 0/30 & 2/30 & 6/30 & 0/30 && - & - & - & - \\
GBDA & 0/30 & 0/30 & 0/30 & 0/30 && 0/10 & 0/10 & 0/10 & 0/10 \\
\bottomrule
\end{tabular}
\label{inv6}
\end{table}

\begin{table}[H]
\centering
\caption{Trigger inversion results for Flan-t5-large on Yahoo data set at high poison rate.}
\begin{tabular}{lcc c cc}
\toprule
& \multicolumn{2}{c}{Tell me seriously(3 tokens)} && \multicolumn{2}{c}{mb(1 token)} \\
\cmidrule(r){2-3} \cmidrule(l){5-6}
& clean label & dirty label && clean label & dirty label \\
\cmidrule(lr){2-2} \cmidrule(lr){3-3} \cmidrule(lr){5-5} \cmidrule(lr){6-6}
CSO(discrete) & 27/30 & 30/30 && 10/10 & 10/10 \\
MM & 11/30 & 15/30 && 10/10 & 10/10 \\
ALT & 18/30 & 17/30 && 10/10 & 10/10 \\
DBS & 3/30 & 0/30 && 5/10 & 2/10 \\
UAT & 2/30 & 2/30 && 0/10 & 0/10 \\
PICCOLO & 0/30 & 0/30 && - & - \\
GBDA & 0/30 & 1/30 && 0/10 & 0/10 \\
\bottomrule
\end{tabular}

\label{inv7}
\end{table}

\end{document}